# Global structure of integer partition sequences

N. M. Chase, Ph.D.
School of Arts and Sciences
Massachusetts College of Pharmacy
 and Health Sciences
179 Longwood Avenue
Boston, MA 02115   U.S.A.

nchase@mcp.edu

## Abstract

Integer partitions are deeply related to many phenomena in statistical physics. A question naturally arises which is of interest to physics both on "purely" theoretical and on practical, computational grounds. Is it possible to apprehend the *global* pattern underlying integer partition sequences and to express the global pattern compactly, in the form of a "matrix" giving all of the partitions of N into exactly M parts? This paper demonstrates that the global structure of integer partitions sequences (IPS) is that of a complex tree. By analyzing the structure of this tree, we derive a closed form expression for a map from (N, M) to the set of all partitions of a positive integer N into exactly M positive integer summands without regard to order. The derivation is based on the use of modular arithmetic to solve an isomorphic combinatoric problem, that of describing the global organization of the sequence of all ordered placements of N indistinguishable balls into M distinguishable non-empty bins or boxes. This work has the potential to facilitate computations of important physics and to offer new insights into number theoretic problems.

# 1. Introduction

The integer partitioning problem [1 - 4] is deeply connected to many phenomena in statistical physics . A recent series of ground-breaking papers [5 - 7] brought out profound connections between number theoretic questions and the physics of ideal Bose gases. Reference [7], for example, constructed a map from the microcanonical statistics of a thermally isolated non-interacting Bose gas to the partitions of integers, thereby shining a floodlight not only on physics but also on the asymptotics of the number partitioning distribution. So-called "balls in boxes" models, all directly related to integer partitions, illuminate many areas of physics, ranging from lattice gravity [8, 9] to branched polymers [10, 11]. A fascinating recent work [12], for example, explored the finite size scaling behavior and phase transitions in the canonical ensemble of a balls in boxes model. We also now know that a deep connection exists between multidimensional integer partitions and the Potts model [13, 14].

In short, much important work in physics [1] requires the input of information about integer partition sequences but calculations via recursion relations are notoriously slow. (The total number of partitions of an integer N increases exponentially as the square root of N. [2]) This hinders investigation of even relatively small physical systems. Further, the use of approximations for very large N in the thermodynamic limit, though theoretically reasonable, lacks aesthetic appeal.

A question naturally arises which is of interest to physics both on "purely" theoretical and on practical, computational grounds. Is it possible to apprehend the *global* pattern underlying integer partition sequences and to express the global pattern compactly, in the form of a "matrix" giving all of the partitions of N into exactly M parts? This paper demonstrates that such a matrix is straightforwardly derived by using modular arithmetic to describe the global organization of the sequence of all ordered placements of N indistinguishable balls into M distinguishable non-empty bins or boxes.

---

[1] **The reader is referred to references 5 through 14 and all references therein. Apologies are extended to the many authors whose important and pertinent works have not been cited in this brief introduction.**

Specifically, this paper derives a map from the ordered pair (N, M) to the set of all partitions

$$\{s_0, s_1, \ldots, s_{M-1}\} \quad (1)$$

of a positive integer N into exactly M positive integer parts ordered so that

$$s_0 \geq s_1 \ldots \geq s_{M-1} ; \quad (2)$$

$$\sum_{i=0}^{M-1} s_i = N .$$

For example, as shown in Table 5, N =10 can be partitioned into M = 4 positive integer parts in nine different ways:   10 = 7 + 1 + 1 + 1 = 6 + 2 + 1 + 1 = . . . = 3 + 3 + 2 + 2.

In this case, the integer partitions sequence (IPS) "matrix" would have nine rows and four columns, the first row being {7, 1, 1, 1} and the last row being {3, 3, 2, 2}.   In the general case, the IPS matrix has M columns and a number of rows equal to the number, p(N, M),  of partitions of N into exactly M parts.

The paper is organized as follows.   In Section 2 we provide an overview of the integer partitions problem considered here, present the paper's overall strategy for finding a solution, and show (amongst other things) that the partition sequence matrix is (not surprisingly) defined piecewise, having distinct forms for $M \geq \frac{N}{2}$ and $M < \frac{N}{2}$.   In Section 3 we demonstrate that the global structure of integer partition sequences (IPS) is that of a complex tree, the organization of which is readily apprehended by using modular arithmetic.   Having deciphered the global structure of the tree, we present a closed form expression for the IPS matrix as a function of M and a quantity k =  N - M.

Section 4 integrates the developments of the previous sections and, using more compact notation, presents the piecewise-defined map from (N, M) to the matrix of all partitions of integer N into exactly M parts.   Section 5 comments briefly on the computational usefulness of the matrix and suggests potential directions for future work.   An Appendix provides three examples of partition sequences, illustrating for each the emergence of the pattern apprehended by the IPS map developed in this work.

## 2. Overview of the Problem and Strategy

As already noted, the problem considered in this paper is isomorphic to that of apprehending the global structure of the "matrix" enumerating the set of all ordered placements of N indistinguishable balls into M non-empty distinguishable bins. Each row of the IPS matrix corresponds to a distinct distribution of balls. (1) represents an individual row of the matrix, where the integer $s_i$ gives the occupancy of the i'th bin, subject to (2).

The IPS matrix will be derived by the following strategy. Start with the most "condensed" distribution of N balls and then systematically redistribute the balls, in all possible ways consistent with (2), until the distribution becomes as widely dispersed as possible. Tracking this sequence of redistributions is facilitated by defining the quantity k = N - M. In the most "condensed" distribution of balls over the M non-empty bins, we clearly have (k + 1) balls in bin 0 and one ball in each of the other bins.

| $s_0$ | $s_1$ | $s_2$ | | $s_{M-1}$ |
|---|---|---|---|---|
| **k + 1** | **1** | **1** | . . . | **1** |
| bin 0 | bin 1 | bin 2 | | bin (M - 1) |

**Table 1: Initial distribution of N = M + k balls in M non-empty boxes; this is the first row of the IPS matrix.**

That is,

$$s_i = \begin{cases} k+1, & i = 0 \\ 1, & 1 \leq i \leq M-1 \end{cases} \tag{3}$$

Note, for future reference, that the zero'th bin has k "excess balls" above the singlet occupancy of all others.

The IPS matrix is clearly piecewise defined, having distinct structures for $M < k$ and $M \geq k$. If $M < k$, then in the most widely "dispersed" state of the system, each of the M bins will have occupancy greater than one. On the other hand, if $M \geq k$, then in the most widely dispersed distribution, k bins will have occupancy greater than 1 and M - k bins will have singlet occupancy (so at least one bin will have singlet occupancy).

**Partitions for** $M \geq \dfrac{N}{2}$   ($M \geq k$)

Let us consider the various ways in which the "k excess balls" originally in bin 0 may be redistributed over M non-empty distinguishable bins, subject to (2). We introduce the symbol j[i]; within an individual row of the IPS matrix, the positive integer j[i] will specify the number of balls which have been shifted from the 0'th bin to the i'th bin. Since we conserve the number of balls ($N = s_0 + s_1 + ... + s_{M-1}$), an individual row of the IPS matrix will have the form shown in Table 2.

| $s_0$ | $s_1$ | $s_2$ | | $s_{k-1}$ | | | | $s_{M-1}$ |
|---|---|---|---|---|---|---|---|---|
| $(k+1) - \sum_{h=1}^{k-1} j[h]$ | 1 + j[1] | 1 + j[2] | .. | 1 + j[k - 1] | 1 | 1 | ... | 1 |
| bin 0 | bin 1 | bin 2 | | bin (k - 1) | | | | bin (M - 1) |

**Table 2: The structure of individual rows of the IPS matrix for M greater than or equal to N/2;**

That is,

$$s_i = \begin{cases} (k+1) - \sum_{h=1}^{k-1} j[h], & i = 0 \\ 1 + j[i], & 1 \leq i \leq k - 1 \\ 1, & k \leq i \leq M - 1 \end{cases}. \quad (4)$$

Each row of the IPS matrix will incorporate its own unique sequence of values for the integers j[i].

Ordering of the elements within the individual rows of the IPS matrix according to (2) requires that

$$\left(k - \sum_{h=1}^{k-1} j[h]\right) \geq j[1] \geq j[2] \geq \ldots \geq j[k-1] \quad . \tag{5}$$

Note that since the (k - 1)st bin receives at most 1 of the k "excess balls" originally in bin 0 (c.f. the Hindenburg algorithm [3, 4]), j[k - 1] clearly takes on values of only 0 and 1.

Now condition (5) can be implemented by increasing j[a], for $1 \leq a \leq k - 2$, in unit steps from j[a + 1] up to a certain maximum value, which we will for now refer to as $J_{max}[a]$.

That is, for a given N and M such that $M \geq N/2$, the full IPS matrix will be generated by repeated, recursive evaluation of $\left\{\left(k + 1 - \sum_{h=1}^{k-1} j[h]\right), (1 + j[1]), (1 + j[2]), \ldots, (1 + j[k-1]), 1, \ldots\right\}$ for integer j[k -1] ranging from 0 to 1, then for j[k - 2] ranging in unit steps from j[k - 1] to a certain maximum Jmax[k - 2], then for j[k - 3] ranging in unit steps from j[k - 2] to a certain maximum Jmax[k - 3], . . . ,

and finally j[1] ranging in unit steps from j[2] to a certain maximum Jmax[1].

One can already see that the global structure of integer partition sequences is encoded in a complex tree of j values. In Section 3, we will examine the tree's global structure and thereby derive an expression for Jmax[a]. At this early point it is, however, worth noting that conditions (2) and (5) will require that the maximum values of the integers j[a], for $1 \leq a \leq k - 2$, depend upon the set of all integers j[h] for h ranging from a + 1 to k - 1. A more detailed notation for the quantity we have referred to as Jmax[a] is thus $J_{max}^{(+)}\left[a, \{j[h]\}_{h=a+1}^{k-1}\right]$ for $1 \leq a \leq k - 2$, where the superscript (+) denotes partitions in which $M \geq k$.

**Partitions for** $M < \dfrac{N}{2}$   $(M < k)$

Again, with k defined as N - M, we begin with the most condensed state, given in Table 1 and repeated here for convenience.

| $s_0$ | $s_1$ | $s_2$ | | $s_{M-1}$ |
|---|---|---|---|---|
| **k + 1** | **1** | **1** | . . . | **1** |
| bin 0 | bin 1 | bin 2 | | bin (M - 1) |

**Table 3:   Initial distribution of N = M + k balls in M non-empty boxes; this is the first row of the IPS matrix.**

We again seek the set of all possible redistributions of the "k excess balls" originally in bin 0 into the other bins, subject to (2); but now, since N exceeds 2M, every bin will contain at least 2 balls in the system's most dispersed state. Table 4 shows the structure of individual rows of the IPS matrix for M < k, where again the positive integers j[i] give the number of balls which have been shifted from bin 0 to the i'th bin.

| $s_0$ | $s_1$ | $s_2$ | | $s_{M-1}$ |
|---|---|---|---|---|
| $(k+1) - \sum_{h=1}^{M-1} j[h]$ | **1 + j[1]** | **1 + j[2]** | . . . | **1 + j[M - 1]** |
| bin 0 | bin 1 | bin 2 | | bin (M - 1) |

**Table 4:   The structure of individual rows of the IPS matrix for M less than N/2.**

That is,

$$s_i = \begin{cases} (k+1) - \sum_{h=1}^{M-1} j[h]\,, & i = 0 \\ 1 + j[i]\,, & 1 \leq i \leq M-1 \end{cases}. \qquad (6)$$

Again, of course, each individual row of the IPS matrix will incorporate its own unique sequence of values for the integers j[i].

Ordering according to (2) now requires that

$$\left(k - \sum_{h=1}^{M-1} j[h]\right) \geq j[1] \geq j[2] \geq \ldots \geq j[M-1] \quad . \tag{7}$$

Note that, not surprisingly, the structure of the leftmost term of (7) is almost identical to the corresponding term of (5), the only difference being that in (7) the upper limit of the sum must be (M - 1) (and not k - 1) in order to conserve the number of balls during redistribution over at most M (not k) cells.

Now condition (7) will be imposed by increasing the integers j[a], for $1 \leq a \leq M - 2$, in unit steps from j[a + 1] up to a certain maximum value which we will, for the moment, refer to as $J_{max}[a]$ for $1 \leq a \leq M - 2$.

For partitions in which M < k, the full IPS matrix will be generated by repeated, recursive evaluation of

$$\left\{\left(k + 1 - \sum_{h=1}^{M-1} j[h]\right), (1 + j[1]), (1 + j[2]), \ldots, (1 + j[M-1])\right\}$$

for j[M -1] ranging in unit steps from 0 to Jmax[M - 1],

then for j[M - 2] ranging in unit steps from j[M - 1] to a certain maximum Jmax[M - 2],

then for j[M - 3] ranging in unit steps from j[M - 2] to a certain maximum Jmax[M - 3], . . . ,

and finally j[1] ranging in unit steps from j[2] to a certain maximum Jmax[1].

Conditions (2) and (7) will require that the maximum values of the integers j[a], for $1 \leq a \leq M - 2$, depend upon the set of all integers j[h] from h = a + 1 to M - 1.

A more detailed notation for the quantity we have referred to as Jmax[a] is thus $J_{max}^{(-)}\left[a, \{j[h]\}_{h=a+1}^{M-1}\right]$

for $1 \leq a \leq M - 2$, where the superscript ( - ) denotes partitions in which M < k.

The maximum occupancy of the (M - 1)st cell, and thus the maximum value of j[M - 1], is straightforwardly derived as follows. In the most dispersed state of the system, N balls are distributed as equally as possible over all M non-empty cells, subject to (2). Thus, in the most dispersed state, each of the M bins will have occupancy at least equal to the integer part of N/M, which is given by

$$\frac{N - Mod[N, M]}{M}. \tag{8a}$$

Now we may imagine constructing the most dispersed configuration by first filling all bins equally up to the value given in (8a).  We would then still have Mod[N,M] balls left over to be sprinkled into bins as evenly as possible, subject to condition (2).   However,  the (M - 1)st bin is surely exempt from this sprinkling, since $M < \frac{N}{2}$.   Thus, the maximum *occupancy* of bin (M -1) is given by (8a).  Since $s_{M-1} = 1 + j[M-1]$, the greatest value of j[M - 1] is given by

$$J^{(-)}_{max}[M-1] = \left( \frac{N - Mod[N, M]}{M} \right) - 1 \, . \tag{8b}$$

## 3.    Global Structure of the IPS Matrix Tree of j Values

We begin this section by considering the structure of a relatively small partition sequence with M < k,  examining its tree of j values as generated by the familiar Hindenburg algorithm [3, 4]. Now the Hindenburg algorithm begins by incrementing values of $s_0$ and $s_1$; it specifies only "localized" structures within the tree and is executed independent of the "large scale" organization of the full sequence of partitions being sought.   Although Section 2 has already provided evidence that the *global* structure of the tree is most readily be mapped by starting at the "roots" of the tree (at the values of the integers j[M - 1]),  a deliberately "backwards" initial approach to the problem serves two purposes. It will provide:  1) easy numerical examples of fairly complex combinatorial arguments to be used in solving the more formidable global problem, and  2) further insight into aspects of the tree structure to be incorporated in the upper limits of the integers j[i].   After this brief examination of localized structures, similar combinatorial arguments will be applied to surmise the global tree structure and to derive general expressions for $J^{(-)}_{max}\left[a, \{j[h]\}_{h=a+1}^{M-1}\right]$ and $J^{(+)}_{max}\left[a, \{j[h]\}_{h=a+1}^{k-1}\right]$ .

## 3A Combinatorics Applied to Localized Aspects of the Tree Structure

For N = 10 balls being distributed into M = 4 bins, Table 5 demonstrates that, as generated by the Hindenburg algorithm, the sequence of partitions emerges in readily discernable *stages*, beginning with the most compact distribution of balls.

| N = 10, M = 4, k = N - M = 6 | | | |
|---|---|---|---|
| $s_0$ | $s_1$ | $s_2$ | $s_3$ |
|  | 1 + j[1] | 1 + j[2] | 1 + j[3] |
| 7 | 1 | 1 | 1 |
| 6 | 2 | 1 | 1 |
| 5 | 3 | 1 | 1 |
| 4 | 4 | 1 | 1 |
| 5 | 2 | 2 | 1 |
| 4 | 3 | 2 | 1 |
| 3 | 3 | 3 | 1 |
| 4 | 2 | 2 | 2 |
| 3 | 3 | 2 | 2 |

**Table 5:** The IPS matrix for N = 10, M = 4.

In the r'th stage, balls are shifted among bins 0 through r, while $s_{r+1}$ through $s_{M-1}$ remain equal to 1. This *sequence of stages* proceeds until, in the most dispersed distribution, the N balls are distributed as equally as possible over all M bins, subject to (2).

Let us examine the first two stages.

<u>In the first stage</u>, $s_0$ is incremented down in unit steps, while $s_1$ is correspondingly incremented upwards. Both j[3] and j[2] remain equal to zero during this stage. The k excess balls (above singlet background) originally in the zero'th bin are shifted one by one until they are finally distributed as equally as possible between bins 0 and 1(redistribution between *two* bins). Thus, in this stage, the maximum value of j[1] is given by

$$J_{max}[1] = \frac{k - Mod[k,2]}{2} = \frac{6 - Mod[6,2]}{2} = 3 \ . \tag{9}$$

In the second stage, $S_0, S_1, S_2$ are adjusted. Only j[3] remains equal to zero during this stage.

Balls are shifted amongst bins 0, 1, and 2 until finally the k excess balls (above singlet background) originally in the zero'th bin are distributed as equally as possible over these *three* bins. Thus, in this stage, j[2] has a greatest value

$$J_{max}[2] = \frac{k - Mod[k,3]}{3} = \frac{6 - Mod[6,3]}{3} = 2 \ . \tag{10}$$

*However, at the same time*, j[1] is incremented in unit steps from j[2] up to a maximum value such that balls not already shifted into bin 2 are finally distributed as equally as possible between bins 0 and 1 (two bins). Thus, in the second stage,

$$J_{max}[1] = \frac{k - j[2] - Mod[k - j[2], 2]}{2} = \frac{6 - j[2] - Mod[6 - j[2], 2]}{2}. \tag{11}$$

Comparison of (9) and (11) demonstrates that $J_{max}[1]$ depends upon input of j[2].

Both (9) and (11) are included in the following expression

$$J_{max}[1, j[2]] = \frac{6 - j[2] - Mod[6 - j[2], 2]}{2} = \begin{cases} 3, & j[2] = 0 \\ 2, & j[2] = 1 \\ 2, & j[2] = 2 \end{cases}. \tag{12}$$

(Recall that in stage 1, j[2] = 0.)

This concludes our brief excursion "backwards" through the tree.

## 3B     Global Structure of Integer Partitions Matrix for $M < k$

The search for a general expression for $J^{(-)}_{max}\left[a, \{j[h]\}_{h=a+1}^{M-1}\right]$ is best initiated at the "roots" of the IPS matrix's tree of j values, at the set of j[M - 1] values, to which we now turn.

We have already seen that the greatest value of $S_{M-1}$ is given by

$$S_{M-1} = \frac{M + k - Mod[M + k, M]}{M};  \qquad (13)$$

so that

$$J^{(-)}_{max}[M - 1] = \frac{M + k - Mod[M + k, M]}{M} - 1. \qquad (14)$$

Now for each of the possible values of the integers j[M - 1], which range from 0 up to the value given in (14), we have a set of possible values for j[M - 2], which range from j[M - 1] up to a certain maximum $J^{(-)}_{max}[M - 2, j[M - 1]]$. We shall now use reasoning similar to that leading to (9) through (11) to obtain find the general form of $J^{(-)}_{max}[M - 2, j[M - 1]]$.

Given that some of the k excess originally in bin 0 have been shifted to bin M - 1, the remaining balls must ultimately be distributed as equally as possible between bin M - 2 and those to its left (distribution over M - 1 bins).
Thus,

$$J^{(-)}_{max}[M - 2, j[M - 1]] = \frac{k - j[M - 1] - Mod[k - j[M - 1], M - 1]}{M - 1}. \qquad (15)$$

Let us take this kind of reasoning one step further. For each combination of values of j[M - 1] and j[M - 2], we have a sequence of possible values for j[M - 3], ranging from j[M - 2] up to a certain maximum value determined as follows. Given the number of the k excess balls originally in bin 0 which have been shifted into bins M - 1 and M - 2, the remaining ones must finally be distributed as equally as possible between bin M - 3 and those to *its* left (distribution over M - 2 bins). Thus,

$$J^{(-)}_{max}\left[M-3,\{j[M-1],j[M-2]\}\right] = \frac{k - j[M-1] - j[M-2] - Mod\left[k - j[M-1] - j[M-2], M-2\right]}{M-2}. \tag{16}$$

Leaping ahead to the integer values of j[2], similar reasoning implies

$$J^{(-)}_{max}\left[2,\{j[h]\}_{h=3}^{M-1}\right] = \frac{k - \sum_{h=3}^{M-1} j[h] - Mod\left[k - \sum_{h=3}^{M-1} j[h], 3\right]}{3}; \tag{17}$$

given the redistribution of some of the k excess balls into bins to the right of bin 2, the remaining ones must ultimately be distributed as equally as possible between it (bin 2), bin 1 and bin 0 (*three* bins altogether).
Similarly,

$$J^{(-)}_{max}\left[1,\{j[h]\}_{h=2}^{M-1}\right] = \frac{k - \sum_{h=2}^{M-1} j[h] - Mod\left[k - \sum_{h=2}^{M-1} j[h], 2\right]}{2}, \tag{18}$$

and in general we have

$$J^{(-)}_{max}\left[a,\{j[h]\}_{h=a+1}^{M-1}\right] = \frac{k - \sum_{h=a+1}^{M-1} j[h] - Mod\left[k - \sum_{h=a+1}^{M-1} j[h], a+1\right]}{a+1}, \quad a < M-1. \tag{19}$$

Thus, for $2 < M < k$ ($M < N/2$), the IPS matrix is given by

$$\left\{\left(k + 1 - \sum_{h=1}^{M-1} j[h]\right), (1 + j[1]), (1 + j[2]), \ldots, (1 + j[M-1])\right\}\bigg|_{j[1]=j[2]}^{J_{\max}^{(-)}[1]} \bigg|_{j[2]=j[3]}^{J_{\max}^{(-)}[2]} \cdots \bigg|_{j[M-1]=0}^{J_{\max}^{(-)}[M-1]}, \quad (20)$$

where $J_{\max}^{(-)}[a]$, $1 \leq a \leq M - 2$, is an abbreviated notation for (19) and $J_{\max}^{(-)}[M-1]$ is given by (14). The super and subscripted vertical bars in (20) are used to indicate that the quantity in curly brackets is to be evaluated repeatedly in the manner described in Section 2:

by incrementing j[M - 1] in unit steps from 0 to $J_{\max}^{(-)}[M-1]$,

then incrementing j[M - 2] in unit steps from j[M - 1] to $J_{\max}^{(-)}[M-2]$, ...,

and finally incrementing j[1] in unit steps from j[2] to $J_{\max}^{(-)}[1]$.

For $M = 1$ and $M = 2$, we obtain the simpler partition structures

$$\begin{array}{ll} \{k + 1\}, & M = 1 \\ \\ \left\{(k + 1 - j[1]), (1 + j[1])\right\}\bigg|_{j[1]=0}^{\frac{M+k-\mathrm{Mod}[M+k,M]}{M}}, & M = 2 \end{array} \quad (21)$$

The reader is now referred to the Appendix, which provides three examples which explicitly demonstrate the emergence of the global pattern given by (20) and (14).

## 3C  Global Structure of Integer Partitions Matrix for $M \geq k$

As already noted in Section 2, the last (M - k) elements in each row of the IPS matrix have values equal to 1. Each row of the IPS matrix now has the form

$$\left\{\left(k+1-\sum_{h=1}^{k-1} j[h]\right), (1+j[1]), (1+j[2]), \ldots, (1+j[k-1]), 1, \ldots\right\}, \qquad (22)$$

where j[k - 1] takes on values of 0 or 1.

The same combinatorial arguments which led to (19) (for partitions in which M < k) may be applied, without elaboration, to obtain the general expression for $J_{\max}^{(+)}\left[a, \{j[h]\}_{h=a+1}^{k-1}\right]$, $a < k - 1$, given in (24); redistributions of the k excess balls originally in bin 0 are now over at most k (rather than M) bins.

For partitions in which $M \geq k$ (and k > 2), the IPS matrix thus has the form

$$\left\{\left(k+1-\sum_{h=1}^{k-1} j[h]\right), (1+j[1]), (1+j[2]), \ldots, (1+j[k-1]), 1, \ldots\right\}\Bigg|_{j[1]=j[2]}^{J_{\max}^{(+)}[1]} \Bigg|_{j[2]=j[3]}^{J_{\max}^{(+)}[2]} \cdots \Bigg|_{j[k-1]=0}^{1}, \qquad (23)$$

where $J_{\max}^{(+)}[a]$ is used as an abbreviated notation for

$$J_{\max}^{(+)}\left[a, \{j[h]\}_{h=a+1}^{k-1}\right] = \frac{k - \sum_{s=a+1}^{k-1} j[s] - \text{Mod}\left[k - \sum_{s=a+1}^{k-1} j[s], a+1\right]}{a+1}, \quad a < k - 1. \qquad (24)$$

For k = 0, 1, and 2, we obtain the simpler partition structures

$$\begin{aligned}
&\{1, 1, \ldots\}, &&k = 0 &&(M \text{ summands equal to 1}) \\
&\{2, 1, 1, \ldots\}, &&k = 1 &&(M - 1 \text{ summands equal to 1}) \\
&\{(3 - j[1]), (1 + j[1]), 1, 1, \ldots\}\big|_{j[1]=0}^{1}, &&k = 2 &&(M - 2 \text{ summands equal to 1})
\end{aligned} \qquad (25)$$

## 4. Partitions of N into Exactly M Parts

In this section, we express the IPS matrix in terms of N and M and introduce the following more compact notation. For integers A and B $(B \neq 0)$, $\frac{A}{B} = \left\lfloor \frac{A}{B} \right\rfloor + \frac{Mod[A,B]}{B}$, where $\left\lfloor \frac{A}{B} \right\rfloor$ is the greatest integer less than or equal to $\frac{A}{B}$.

### IPS Matrix for $M < N/2$

By replacing k by N - M in (20), we obtain

$$\left\{ \left(N - M + 1 - \sum_{h=1}^{M-1} j[h]\right), (1 + j[1]), (1 + j[2]), \ldots, (1 + j[M-1]) \right\} \Bigg|_{j[1]=j[2]}^{J_{max}^{(-)}[1]} \Bigg|_{j[2]=j[3]}^{J_{max}^{(-)}[2]} \cdots \Bigg|_{j[M-1]=0}^{J_{max}^{(-)}[M-1]}, \quad (26)$$

where $J_{max}^{(-)}[M-1]$ is given by (14) and $J_{max}^{(-)}[a]$, $1 \leq a \leq M - 2$, is a shorthand notation for (19). (19) may be written more compactly as

$$J_{max}^{(-)}\left[a, \{j[h]\}_{h=a+1}^{M-1}\right] = \left\lfloor \frac{N - M - \sum_{h=a+1}^{M-1} j[h]}{a + 1} \right\rfloor, \quad a < M - 1, \quad (27)$$

Rewriting (14) in terms of N and M, we obtain

$$J_{max}^{(-)}[M - 1] = \frac{N - Mod[N, M]}{M} - 1 = \left\lfloor \frac{N}{M} \right\rfloor - 1. \quad (28)$$

Replacing k by N - M, we may express the simpler partitions given in (21) as

$$\begin{matrix} \{N - M + 1\}, & M = 1 \\ \\ \left\{(N - M + 1 - j[1]), (1 + j[1])\right\}\Big|_{j[1]=0}^{\lfloor N/M \rfloor}, & M = 2 \end{matrix} \quad (29)$$

## IPS Matrix for $M \geq N/2$

Substituting k = N - M into (23), we obtain

$$\left\{\left(N-M+1-\sum_{h=1}^{N-M-1}j[h]\right),(1+j[1]),(1+j[2]),\ldots,(1+j[N-M-1]),1,\ldots\right\}\Bigg|_{j[1]=j[2]}^{J_{\max}^{(+)}[1]}\Bigg|_{j[2]=j[3]}^{J_{\max}^{(+)}[2]}\cdots\Bigg|_{j[N-M-1]=0}^{1}, \quad (30)$$

where $J_{\max}^{(+)}[a]$ is an abbreviated notation for (24).

Using the same bracket notation we have used in (27) and (28), and replacing k by N - M, (24) may be written more compactly as

$$J_{\max}^{(+)}\left[a,\{j[h]\}_{h=a+1}^{N-M-1}\right]=\left\lfloor\frac{N-M-\sum_{s=a+1}^{N-M-1}j[s]}{a+1}\right\rfloor, \quad a < N-M-1. \quad (31)$$

With substitution of k for N - M into (25), the simpler partitions in which k = 0, 1, and 2 may be expressed as

$$\{1,1,\ldots\}, \qquad\qquad N=M \quad (M\text{ summands equal to 1})$$

$$\{2,1,1,\ldots\}, \qquad\qquad N=M+1 \quad (M-1\text{ summands equal to 1}). \quad (32)$$

$$\{(3-j[1]),(1+j[1]),1,1,\ldots\}\Big|_{j[1]=0}^{1}, \quad N=M+2 \quad (M-2\text{ summands equal to 1})$$

## 5. Concluding Remarks

Given the simplicity of the derivation presented in this paper, it is difficult to believe that the IPS matrix (or a simpler variant of it) is not already well known; however, the author has found it nowhere in the literature, or in Dickson's [2] three volume exposition.

Now computational speed is not, per se, the focus of this paper. Nevertheless, for completeness this section provides preliminary evidence that the IPS matrix is capable of greatly facilitating statistical physics calculations.

Instinct suggests that the greater the amount of "information" embodied in the structure of a mathematical expression, the faster the expression would *in principle be able* to generate its intended numerical output. Significant "processing time" would be saved if much of the required "processing" has already been done in constructing the expression itself. From this viewpoint, we have asked whether the IPS matrix represents an improvement over calculations via a recursion relation.
As a first test, we considered the following well known recurrence relation

$$p[N, M] = p[N-1, M-1] + p[N-M, M], \qquad (33)$$

$$\text{where } p[N, M] = \begin{cases} 0 & \text{for } N < M \\ 0 & \text{for either } N < 1 \text{ or } M < 1 \\ 1 & \text{for } N = M \\ 1 & \text{for } M = 1 \end{cases},$$

which generates the *numbers* of *partitions* of a positive integer N into exactly M positive integer parts. As a compact standard of comparison of (33) with the IPS matrix, we have considered the total number of partitions P[N] of an integer N, which are tabulated in [15]. (P[N] is obtained by summing p[N,M] over M from M = 1 to M = N.)

Preliminary results appear impressive. For example, the recurrence relation (33) predicts the number P[50] = 204,226 in 28.3 seconds, while the IPS matrix, in only 20.6 seconds, computes not only this single numerical value (P[50]) but also the full array of integer partition sequences, all 204,226 of them (for N = 50 partitioned into up to 50 positive integer parts). The IPS matrix produces comparably remarkable results for all N > 40 thus far tested. Now programming techniques are of unquestionable significance in any comparison of speeds. Although the author is a novice programmer in MATHEMATICA, the results presented above nevertheless appear significant: - "output" of (33) was (eventually) obtained by writing (33) as a single simple function (not by solving the recursion relation in closed form). On the other hand, the IPS matrix was coded in terms of the complicated set of recursive evaluations given in Section 4; it is a near certainty that more efficient code could have been written.

We close by suggesting potential further developments. The integer partitioning problem underlies many open questions in mathematics, answers to which would likely inform important issues in physics. The IPS matrix, unlike the Hindenburg algorithm, reflects the *global* structure of integer partition sequences. In this sense, it represents a more "comprehensive" representation of IPS. However, the author suspects that its complex tree-like structure could be expressed much more concisely.

Thus far, attempts to simplify the IPS matrix have been unsuccessful. On the other hand, by counting the number of paths through the IPS matrix tree, the author has succeeded in deriving a closed form expression for p[N,M], the number of partitions of an integer N into exactly M parts. From this,
a simple sum over M from 1 to N produces a transparent alternative to the Hardy-Ramanujan-Rademacher formula for p(N) [2, 3, 16]. These results will be presented in a forthcoming work. We further note that the author has used the global structure of IPS to derive a possibly interesting relation between prime numbers and modular arithmetic forms; long sequences of consecutive prime numbers have been accurately generated based on a "balls in boxes" interpretation of prime numbers.

# APPENDIX

**Table 6:** IPS Matrix Tree of "j values" for N=10, M = 4. (k = N - M). The superscript ( - ) has been omitted from Jmax[a] for convenience.

| N = 10, M = 4, k = 6 | | | | ILLUSTRATION OF THE GLOBAL TREE STRUCTURE GIVEN BY (26), (27), AND (28). | | |
|---|---|---|---|---|---|---|
| | 1 + j[1] | 1 + j[2] | 1 + j[3] | | | |
| 7 | 1 | 1 | 1 | | | |
| 6 | 2 | 1 | 1 | j[3]=0, j[2] = 0 | j[3]=0, j[2] = 0, | |
| 5 | 3 | 1 | 1 | j[1]=0 to 3 | j[1] = j[2] to Jmax[1] | |
| 4 | 4 | 1 | 1 | | j[3] = 0, | |
| | | | | | j[2] = j[3] to 2, | |
| | | | | | j[1] = j[2] to Jmax[1] | |
| 5 | 2 | 2 | 1 | j[3]=0, j[2] = 1 | j[3]=0, j[2] = 1, | j[3] = 0, |
| 4 | 3 | 2 | 1 | j[1]=1 to 2 | j[1] = j[2] to Jmax[1] | j[2] = j[3] to Jmax[2] |
| | | | | | | j[1] = j[2] to Jmax[1] |
| 3 | 3 | 3 | 1 | j[3]=0, j[2] = 2, | | |
| | | | | j[1]=2 to 2 | | |
| | | | | | j[3] = j[2] to Jmax[1] | |
| 4 | 2 | 2 | 2 | j[3]=1, j[2] = 1 | j[3]=1, j[2] = 1, | j[3] = 1, |
| 3 | 3 | 2 | 2 | j[1]=1 to 2 | j[1] = j[2] to Jmax[1] | j[2] = j[3] to Jmax[2] |
| | | | | | | j[1] = j[2] to Jmax[1] |
| | | | | | | j[3] = 0 to Jmax[3] |
| | | | | | | j[2] = j[3] to Jmax[2] |
| | | | | | | j[1] = j[2] to Jmax[1] |

$$J_{\max}[1] = \frac{(6 - \sum_{h=2}^{3} j[h]) - Mod[(6 - \sum_{h=2}^{3} j[h]), 2]}{2} \quad (A1)$$

$$J_{\max}[2] = \frac{(6 - \sum_{h=3}^{3} j[h]) - Mod[(6 - \sum_{h=3}^{3} j[h]), 3]}{3} = \frac{(6 - j[3]) - Mod[(6 - j[h]), 3]}{3} = \begin{cases} 2, & j[3] = 0 \\ 1, & j[3] = 1 \end{cases} \quad (A2)$$

$$J_{\max}[3] = \frac{M + k - Mod[(M+k), M]}{M} - 1 = \frac{10 - Mod[10,4]}{4} - 1 = 1 \quad (A3)$$

**Table 7:** IPS Matrix Tree of "j values" for N=12, M = 5. (k = N - M). The superscript ( - ) has been omitted from Jmax[a] for convenience.

| N = 12, M = 5, k =7 | | | | | ILLUSTRATION OF THE GLOBAL TREE STRUCTURE GIVEN BY (26), (27), AND (28). | | | |
|---|---|---|---|---|---|---|---|---|
| | 1+j[1] | 1+j[2] | 1+j[3] | 1+j[4] | | | | |
| 8 | 1 | 1 | 1 | 1 | | | | |
| 7 | 2 | 1 | 1 | 1 | j[4] = 0, | j[4] = 0, | | |
| | | | | | j[3]=0, j[2] = 0, | | | |
| 6 | 3 | 1 | 1 | 1 | j[1] =0 to 3 | j[2] = 0, | | |
| | | | | | | j[1] = j[2] to Jmax[1] | | |
| 5 | 4 | 1 | 1 | 1 | | | | |
| 6 | 2 | 2 | 1 | 1 | j[4] = 0, j[3] = 0, | j[4] = 0, j[3] = 0, | | |
| | | | | | j[2] = 1, | | | |
| 5 | 3 | 2 | 1 | 1 | j[1]= 1 to 3 | j[2] = 1, | j[4] = 0, | |
| | | | | | | j[1] = j[2] to Jmax[1] | j[3] = j[4] to Jmax[3] | j[4] = 0 to Jmax[4] |
| 4 | 4 | 2 | 1 | 1 | | | j[2] = j[3] to Jmax[2] | j[3] = j[4] to Jmax[3] |
| | | | | | | | j[1] = j[2] to Jmax[1] | j[2] = j[3] to Jmax[2] |
| 4 | 3 | 3 | 1 | 1 | j[4] = 0, j[3] = 0, | j[4] = 0, j[3]=0, | | j[1] = j[2] to Jmax[1] |
| | | | | | j[2] = 2, | j[2] = j[3] to 2 | | |
| | | | | | j[1] = j[2] to Jmax[1] | j[1] = j[2] to Jmax[1] | | |
| 5 | 2 | 2 | 2 | 1 | j[4] = 0, j[3] = 1, | j[4] = 0, j[3]=1, | | |
| | | | | | j[2] = 1, | j[2] = j[3] to 2 | | |
| | | | | | j[1] = 1 to 2 | j[1] = j[2] to Jmax[1] | | |
| 4 | 3 | 2 | 2 | 1 | j[4] = 0, j[3] = 1, | j[4] = 0, j[3]=1, | | |
| | | | | | j[2] = 1, | j[2] = j[3] to Jmax[2] | | |
| | | | | | j[1] = j[2] to Jmax[1] | j[1] = j[2] to Jmax[1] | | |
| 3 | 3 | 3 | 2 | 1 | j[4] = 0, j[3] = 1, | j[4] = 0, j[3]=1, | | |
| | | | | | j[2] = 2, | j[2] = j[3] to Jmax[2] | | |
| | | | | | j[1] = j[2] to Jmax[1] | j[1] = j[2] to Jmax[1] | | |
| 4 | 2 | 2 | 2 | 2 | j[4] = 1, j[3] = 1, | j[4] = 1, j[3] = 1, | j[4] = 1, | |
| | | | | | j[2] = 1, | j[2] = j[3] to 1 | j[3] = j[4] to Jmax[3] | |
| | | | | | j[1] = 1 to 2 | j[1] = j[2] to Jmax[1] | j[2] = j[3] to Jmax[2] | |
| | | | | | | | j[1] = j[2] to Jmax[1] | |
| 3 | 3 | 2 | 2 | 2 | j[4] = 1, j[3] = 1, | j[4] = 1, j[3] = 1, | | |
| | | | | | j[2] = 1, | j[2] = j[3] to Jmax[2] | | |
| | | | | | j[1] = j[2] to Jmax[1] | j[1] = j[2] to Jmax[1] | | |
| – | | | | | | | | |

$$J_{\max}[1] = \frac{(7 - \sum_{h=2}^{4} j[h]) - Mod[(7 - \sum_{h=2}^{4} j[h]), 2]}{2} \quad (A4)$$

$$J_{\max}[2] = \frac{(7 - \sum_{h=3}^{4} j[h]) - Mod[(7 - \sum_{h=3}^{4} j[h]), 3]}{3} \quad (A5)$$

$$J_{\max}[3] = \frac{(7 - j[4]) - Mod[(7 - j[4]), 4]}{4} = \begin{cases} 1, j[4] = 0 \\ 1, j[4] = 1 \end{cases} \quad (A6)$$

$$J_{\max}[4] = \frac{M + k - Mod[(M + k), M]}{M} - 1 = \frac{12 - Mod[12,5]}{5} - 1 = 1 \quad (A7)$$

**Table 8: IPS Matrix Tree of "j values" for N=14, M = 6 (k = N - M).** The superscript ( - ) has been omitted from Jmax[a] for convenience.

| N = 14, M = 6, k = 8 | | | | | | ILLUSTRATION OF THE GLOBAL TREE STRUCTURE GIVEN BY (26), (27), AND (28). | | | |
|---|---|---|---|---|---|---|---|---|---|
| | 1+ j[1] | 1+j[2] | 1+j[3] | 1+j[4] | 1 + j[5] | | | | |
| 9 | 1 | 1 | 1 | 1 | 1 | j[5] = j[4] = j[3] = j[2] = 0, | | | j[5] = 0 to Jmax[5] |
| 8 | 2 | 1 | 1 | 1 | 1 | | | | j[4] = j[5] to Jmax[4] |
| 7 | 3 | 1 | 1 | 1 | 1 | j[1] = j[2] to 4 | j[5] = j[4] = j[3] = 0, | | j[3] = j[4] to Jmax[3] |
| 6 | 4 | 1 | 1 | 1 | 1 | | j[2] = j[3] to Jmax[2] | j[5] = j[4] = 0, | j[2] = j[3] to Jmax[2] |
| 5 | 5 | 1 | 1 | 1 | 1 | | j[1] = j[2] to Jmax[1] | j[3] = j[4] to Jmax[3] | j[1] = j[2] to Jmax[1] |
| 7 | 2 | 2 | 1 | 1 | 1 | j[5] = j[4] = j[3] = 0, | | j[2] = j[3] to Jmax[2] | |
| 6 | 3 | 2 | 1 | 1 | 1 | j[2] = 1, j[1] = j[2] to 3 | | j[1] = j[2] to Jmax[1] | |
| 5 | 4 | 2 | 1 | 1 | 1 | | | | |
| 5 | 3 | 3 | 1 | 1 | 1 | j[5] = j[4] = j[3] = 0, | | | |
| 4 | 4 | 3 | 1 | 1 | 1 | j[2] = 2, j[1] = j[2] to 3 | | | |
| 6 | 2 | 2 | 2 | 1 | 1 | j[5] = j[4] = 0, j[3] = 1, | j[5] = j[4] = 0, j[3] = 1 | | |
| 5 | 3 | 2 | 2 | 1 | 1 | j[2] = 1, j[1] = j[2] to 3 | j[2] = 1 | | |
| 4 | 4 | 2 | 2 | 1 | 1 | | j[1] = j[2] to Jmax[1] | | |
| 4 | 3 | 3 | 2 | 1 | 1 | j[5] = j[4] = 0, j[3] = 1, | j[5] = j[4] = 0, j[3] = 1 | | |
| | | | | | | j[2] = 2, j[1] = j[2] to 2 | j[2] = 2 | | |
| 3 | 3 | 3 | 3 | 1 | 1 | j[5] = j[4] = 0, j[3] = 2, | j[5] = j[4] = 0, j[3] = 2 | | |
| | | | | | | j[2] = 2, j[1] = j[2] to 2 | j[2] = 2 | | |
| | | | | | | | j[1] = j[2] to Jmax[1] | | |
| 5 | 2 | 2 | 2 | 2 | 1 | j[5] = 0, j[4] = 1, j[3] = 1, | j[5] = 0, j[4] = 1, j[3] = 1 | j[5] = 0, j[4] = 1, j[3] = 1 | |
| 4 | 3 | 2 | 2 | 2 | 1 | j[2] = 1, j[1] = j[2] to 2 | j[2] = 1 | j[2] = j[3] to Jmax[2] | |
| | | | | | | | j[1] = j[2] to Jmax[1] | j[1] = j[2] to Jmax[1] | |
| 3 | 3 | 3 | 2 | 2 | 1 | j[5] = 0, j[4] = 1, j[3] = 1, | j[5] = 0, j[4] = 1, j[3] = 1 | | |
| | | | | | | j[2] = 2, j[1] = j[2] to 2 | j[2] = 2 | | |
| | | | | | | | j[1] = j[2] to Jmax[1] | | |
| 4 | 2 | 2 | 2 | 2 | 2 | j[5] = 1, j[4] = 1, j[3] = 1, | j[5] = 1, j[4] = 1, j[3] = 1, | j[5] = 1, j[4] = 1, j[3] = 1 | |
| 3 | 3 | 2 | 2 | 2 | 2 | j[2] = 1, j[1] = j[2] to 2 | j[2] = 1, | j[2] = j[3] to Jmax[2] | |
| | | | | | | | j[1] = j[2] to Jmax[1] | j[1] = j[2] to Jmax[1] | |

$$J_{\max}[5] = \frac{M+k - Mod[M+k, M]}{M} - 1 = \frac{14 - Mod[14, 6]}{6} - 1 = 1 \quad (A8)$$

$$J_{\max}[a] = \frac{1}{(a+1)} \left( \left( 8 - \sum_{s=a+1}^{M-1} j[s] \right) - Mod\left[ \left( 8 - \sum_{s=a+1}^{M-1} j[s] \right), (a+1) \right] \right), \ 1 \le a < 5 \quad (A9)$$

25